\documentclass[page-classic]{epl-style-arxiv}

\newcommand{\NJ}[1]{\textcolor{blue}{#1}}

\usepackage{amsmath}
\usepackage{amssymb}
\usepackage{xcolor}
\usepackage{graphicx}
\usepackage{comment}
\usepackage{algorithm}
\usepackage[noend]{algpseudocode}
\usepackage[normalem]{ulem}
\usepackage{here}
\usepackage{hyperref}

\title{Perspective on Physical Interpretations of Rényi Entropy in Statistical Mechanics}
\shorttitle{Title} %Insert here a short version of the title if it exceeds 70 characters

\author{Misaki Ozawa\inst{1} and Nina Javerzat\inst{1}  }%\and S. Author\inst{1} \and T. Author\inst{2}}
\shortauthor{M. Ozawa and N. Javerzat}

\institute{                    
  \inst{1} Univ. Grenoble Alpes, CNRS, LIPhy, 38000 Grenoble, France\\
}

\abstract{
Rényi entropy is a one-parameter generalization of Shannon entropy, which has been used in various fields of physics. Despite its wide applicability, the physical interpretations of the Rényi entropy are not widely known. In this paper, we discuss some basic properties of the Rényi entropy relevant to physics, in particular statistical mechanics, and its physical interpretations using free energy, replicas, work, and large deviation.}

\begin{document}

\maketitle

\section{Introduction}

Entropy plays a central role in statistical descriptions in physics~\cite{sethna2021statistical}. For example, it indicates the direction of time evolution of a macroscopic system.
It can characterize the emergence of various orders in phase transitions and pattern formations. A decrease in entropy suggests the growth of a certain order, and a jump or a kink in entropy implies the existence of a phase transition.
In the information-theoretical perspective, entropy is a measure of uncertainty or of the size of possibility~\cite{cover1999elements}. Thermodynamic entropy in statistical mechanics and the von Neumann entropy in quantum mechanical systems correspond to the Shannon entropy~\cite{shannon1948mathematical}.
%However, measuring the Shannon entropy in physical systems often involves difficulties at many levels, be it in experiments, numerical simulations, or analytical calculations.
Yet, experimental and numerical measurements of the Shannon entropy, as well as analytical calculations are often faced with severe difficulties.
The Rényi entropy is a one-parameter generalization of the Shannon entropy, which was derived in the context of information theory~\cite{Renyi1961measures}. Various other generalizations have been proposed~\cite{Renyi1961measures,tsallis1988possible} (see Refs.~\cite{beck2009generalised,amigo2018brief,ilic2021overview} for review).
%\sout{Notably, the Rényi entropy has a functional form that is suitable for measurements, compared with the Shannon one}
Most notably, the Rényi entropy turns out to be generally much easier to measure than the Shannon entropy, and has thus been used in various fields of physics as an alternative to the latter.
%\NJ{this could be commented a bit more: people who dont know about Renyi might not get it. Or this sentence could be moved after we write explicitly the expression for Rényi (after eq.4 ?}. 
%\MO{Notably, the Rényi entropy is much easier to measure than the Shannon entropy. (We wish to emphasize why we want to use Renyi here. In the end, it is easier to measure. This is the main motivation in many fields. Please edit freely while keeping this message.)}

In quantum-many-body systems for instance % Yet, experimental and numerical measurements, as well as analytical calculations are often faced with severe difficulties. 
the Rényi entropy gives a measure of quantum entanglement. It can be measured in experiments~\cite{Islam2015,kaufman2016quantum} and in numerical simulations that are less computationally demanding~\cite{iaconis2013detecting,alba2016classical,alba2017out,d2020entanglement,zhao2022measuring}. The use of the replica trick and conformal field theory allow moreover to compute analytically Rényi entanglement entropy for some systems, and to obtain the (Shannon or Von Neumann) entanglement entropy by analytical continuation~\cite{Calabrese_2009}.
%On the theoretical side, the Rényi entropy is related to the replica trick, which allows one to compute it analytically for some systems using conformal field theory~\cite{Calabrese_2009}. 
In dynamical systems, it is common practice to use the Rényi entropy as a characterization of chaos~\cite{grassberger1983estimation,halsey1986fractal}.
The Rényi entropy is also related to a theoretical formalism in non-equilibrium dynamics. It is a generalization of the Kolmogorov-Sinai entropy, which corresponds to the cumulant generating function associated with a dynamical partition function involving counting
trajectories in phase space~\cite{beck1995thermodynamics,lecomte2007thermodynamic}.
Since the Rényi entropy is tightly related to the replica trick~\cite{mezard1987spin,charbonneau2023spin}, it has relationships with the physics of disordered systems. Reference~\cite{kurchan2010order} connects the configurational entropy characterizing a thermodynamic glass transition~\cite{cammarota2023kauzmann,berthier2019configurational} and real-space structure through the Rényi entropy~\cite{kurchan2024}.
For other applications in physics, see Ref.~\cite{fuentes2022Renyi} and references therein.

Despite its broad applicability, the physical interpretations of the Rényi entropy have yet to be widely recognized.
In this paper, we review some of these, particularly in the standard statistical mechanics setting.
We first derive the Rényi entropy and summarize general properties without going into mathematical details.
We then discuss its physical interpretations using the Gibbs-Boltzmann distribution. 
%We also discuss the connection with a replica approach in glassy disordered systems. 
Finally, we conclude and discuss perspectives.

\section{Derivation of Rényi entropy}

We review the derivation of the Rényi entropy and some of its general properties that are useful for physical discussions in the next sections.

We consider a probability distribution, ${\bf p}=(p_1, p_2, \cdots, p_\Omega)$ with $p_i \geq 0$ and $\sum_i p_i=1$, where $\Omega$ is the number of events.
To begin with, the Shannon entropy~\cite{shannon1948mathematical}, $S^{\rm Shannon}$, is defined by
\begin{equation}
    S^{\rm Shannon} = - \sum_i p_i \log p_i = \sum_i p_i I_i ,
    \label{eq:def_Shannon}
\end{equation}
where $I_i= - \log p_i$ is the information content or magnitude of surprise (in this paper, the natural logarithm is used unless otherwise stated).
Thus, the Shannon entropy quantifies a {\it mean} or an average of the information content $I_i$. This is a central aspect that motivates the derivation of the Rényi entropy as follows.

In general, the {\it mean} value can be evaluated not only by the standard arithmetic mean (linear average) used in Eq.~(\ref{eq:def_Shannon}) but also by various other types of {\it mean}, such as the geometric mean, harmonic mean, and root mean square (non-linear averages). Recall, for example, that the harmonic mean is used to compute the equivalent resistance of parallel electrical circuits, and the root mean square is widely used in statistical analysis. The concept of {\it mean} can be generalized further~\cite{kolmogorov1930notion,nagumo1930klasse}. One can then define a more general measure of averaged information content, say $S_f$, given by
\begin{equation}
    S_f = f^{-1}\left( \sum_i p_i f(I_i) \right) ,
    \label{eq:mean_general}
\end{equation}
where $f(x)$ is a strictly monotonic and continuous function which has the inverse $f^{-1}(x)$~\cite{kolmogorov1930notion,nagumo1930klasse}.
For example, $f(x)=x$ corresponds to the arithmetic mean, which gives back the  Shannon entropy in Eq.~(\ref{eq:def_Shannon}).
In other examples, $f(x)= \log x$, $f(x)=x^{-1}$, and $f(x)=x^2$ correspond to the geometric mean,  harmonic mean, and root mean square, respectively, which all can be used to evaluate a {\it mean} value of the information content $I_i$.

However, as a quantity of information, one wishes to have an entropy with the property of {\it additivity for independent events} (or extensivity), which is a fundamental property (or a condition) in information theory~\cite{beck2009generalised}.
If two random variables $A$ and $B$ are independent, an entropy $S(A,B)$ of their joint distribution $p_{ij}^{(A,B)}=p_i^{(A)}p_j^{(B)}$ is the sum of their individual entropies:
\begin{equation}
    S(A,B) = S(A) + S(B).
    \label{eq:additivity}
\end{equation}
The additivity (extensivity) is also naturally expected in thermodynamic entropy in physical systems~\footnote{This may not be true for systems with long-range correlations shown in some non-equilibrium systems and systems with long-range interactions~\cite{campa2014physics,mori2013nonadditivity}.}.
The Shannon entropy, which corresponds to the arithmetic mean via $f(x)=x$, satisfies the additivity. Yet, one can easily check that the other means mentioned above, $f(x)= \log x$, $f(x)=x^{-1}$, and $f(x)=x^2$, violate the additivity condition.
Note that other well-known generalized entropies, such as the Tsallis entropy, do not satisfy additivity~\cite{tsallis1988possible,amigo2018brief,ilic2021overview}.

Alfréd Rényi searched for generalized entropies such as Eq.~(\ref{eq:mean_general}) while keeping the additivity condition~\cite{Renyi1961measures}.
He proposed two possible functional forms of $f(x)$.
The first option is $f(x)=a x + b$, where $a$ and $b$ are some constants.
This corresponds to the Shannon entropy in Eq.~(\ref{eq:def_Shannon}).
The second one is $f(x)=c e^{-(q-1)x}$, where $c$ is a constant and $q$ is a parameter (we set $c=1$ without loss of generality). It corresponds to the so-called Rényi entropy 
$S_q^{\rm R\acute{e}nyi}$, given by
\begin{equation}
    S_q^{\rm R\acute{e}nyi} = \frac{1}{1-q} \log \sum_i (p_i)^q ,
    \label{eq:def_Renyi}
\end{equation}
where $q$ is an index with $0<q<\infty$ and $q\neq1$.
$S_q^{\rm R\acute{e}nyi}$ for $q \to 0$, $q \to \infty$, and $q \to 1$ are defined as the corresponding limits $S_a^{\rm R\acute{e}nyi} = \lim_{q \to a} S_q^{\rm R\acute{e}nyi}$ (see below).
By construction, the Rényi entropy quantifies a {\it mean} of information content in a non-linear way, while keeping the additivity for independent events~\footnote{ In terms of Khinchin's axiomatic formulation~\cite{khinchin2013mathematical},
the (Shannon) entropy of a composite system should be the sum of the individual entropies of its components, taking into account conditional dependence and independence. The Rényi entropy corresponds to the less strict condition of additivity in Eq.~(\ref{eq:additivity}), that is, to considering only independent cases. Consequently, the Rényi entropy loses some properties that the Shannon entropy has, such as concavity in the probability distribution (see pedagogical discussions in Refs.~\cite{beck2009generalised,ilic2021overview}).}
%the Rényi entropy corresponds to replacing Axiom 4 by the less strict condition (which is the additivity in Eq.~(\ref{eq:additivity})).
%\MO{Axiom 4 states that the entropy of a composite system should be the sum of the individual entropies of its components, including considerations of conditional dependence and independence. Considering only independent cases gives rise to the Rényi entropy.}
%Consequently, the Rényi entropy loses some properties that the Shannon entropy has, such as concavity in the probability distribution \MO{(see more pedagogical discussions in Refs.~\cite{beck2009generalised,ilic2021overview})}.}
~\cite{Renyi1961measures}.
The additivity in Eq.~(\ref{eq:additivity}) can be easily checked by using the joint distribution $p_{ij}^{(A,B)}=p_i^{(A)}p_j^{(B)}$ for two independent random variables $A$ and $B$:
\begin{eqnarray}
    S_q^{\rm R\acute{e}nyi}(A,B) &=& \frac{1}{1-q} \log \sum_{i,j} \left(p_{ij}^{(A,B)} \right)^q \nonumber \\
&=& \frac{1}{1-q} \log \sum_{i} \left(p_{i}^{(A)} \right)^q + \frac{1}{1-q} \log \sum_{j} \left(p_{j}^{(B)} \right)^q  \nonumber \\
&=& S_q^{\rm R\acute{e}nyi}(A) + S_q^{\rm R\acute{e}nyi}(B).
\end{eqnarray}
As shown in the introduction, the Rényi entropy with various values of $q$ has been widely used in physics, such as a measure of entanglement in quantum systems (e.g., $q=2$ ~\cite{Islam2015,kaufman2016quantum}), in non-equilibrium statistical mechanics ($q=1/2$ and $q \to \infty$)~\cite{dahlsten2011inadequacy,halpern2015introducing}, multi-fractal analysis~\cite{jizba2004world}, characterization of chaos ($q=2$)~\cite{grassberger1983estimation}, and as an estimate of the Shannon entropy (from $q=2, 3, ...$)~\cite{zyczkowski2003Renyi}.
One of the main goals of this paper is to provide concise interpretations of the Rényi entropy with index $q$ in physical systems.
\section{General properties of Rényi entropies}

%\sout{We discuss general properties of the Rényi entropy.}\NJ{maybe this sentence is not needed}\MO{ok}

For the uniform distribution, $p_i=1/\Omega$ for all $i$, one obtains $S_q^{\rm R\acute{e}nyi}= \log \Omega$ irrespective of $q$.

\begin{figure}
\begin{center}
\includegraphics[width=0.8\linewidth]{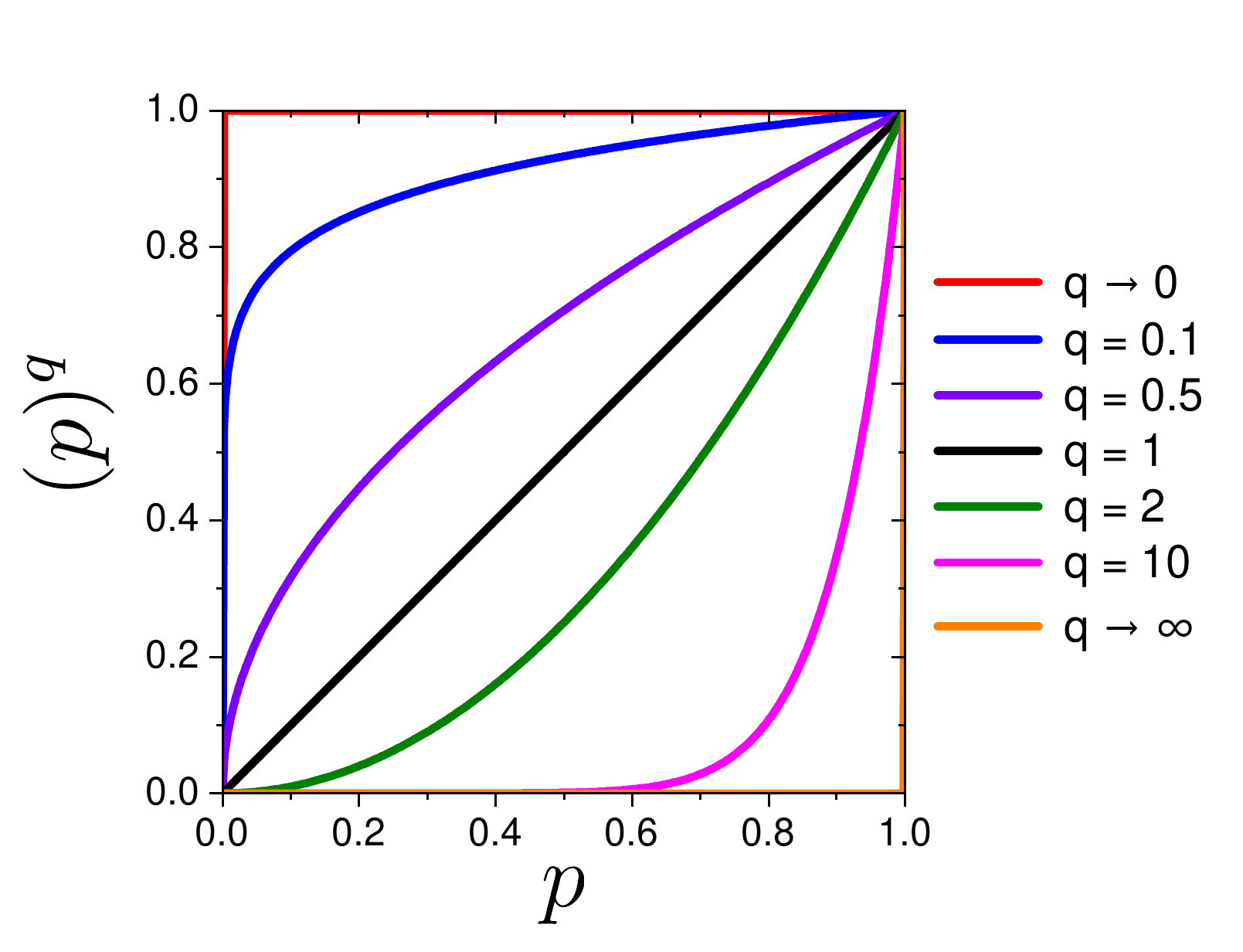}    
\end{center}
\caption{$(p)^q$ for several values of $0<q<\infty$. A larger value of $q$ biases the probability, while a smaller value of $q$ unbiases it.}
\label{fig:p_q}
\end{figure}

Below, we will discuss $S_q^{\rm R\acute{e}nyi}$ for representative values (or limits) of $q$.
We will see that the Rényi entropy unifies various known entropies by the single parameter $q$.

\vspace{0.5cm}
\noindent
i) $q \to 1$ ({\it Shannon entropy}): 

This limit corresponds to the Shannon entropy $S^{\rm Shannon}$ in Eq.~(\ref{eq:def_Shannon}) as easily checked by the l'Hôpital's rule:
\begin{equation}
    S_1^{\rm R\acute{e}nyi} =  - \sum_i p_i \log p_i = S^{\rm Shannon} ,
    \label{eq:Renyi_meets_Shannon}
\end{equation}
%where $S_1^{\rm R\acute{e}nyi} = \lim_{q \to 1} S_q^{\rm R\acute{e}nyi}$. Similarly, we will use a simplified notation, $S_a^{\rm R\acute{e}nyi} = \lim_{q \to a} S_q^{\rm R\acute{e}nyi}$ in this paper.
Equation~(\ref{eq:Renyi_meets_Shannon}) shows that the Rényi entropy is a one parameter generalization of the Shannon entropy.

\vspace{0.5cm}
\noindent
ii) $q \to 0$ ({\it Hartley entropy, Max-entropy}): 

In this limit, only the events with a finite probability, $p_i>0$, contribute to the summation, $\sum_i (p_i)^q$, in Eq.~(\ref{eq:def_Renyi}). Because $(p_i)^q = 1$ for $p_i>0$ and $(p_i)^q = 0$ for $p_i=0$ when $q \to 0$, as shown in Fig.~\ref{fig:p_q}.
Thus, one obtains
\begin{equation}
S_0^{\rm R\acute{e}nyi} = \log \Omega' ,
\label{eq:Max_entropy}
\end{equation}
where $\Omega' \ (\leq \Omega)$ is the number of events with $p_i>0$ (or the size of the support set). 
$S_0^{\rm R\acute{e}nyi}$ characterizes a magnitude of uncertainty by taking into account all the events with $p_i>0$ on equal footing. This is often called the Hartley entropy or max-entropy.

\vspace{0.5cm}
\noindent
iii) $q=2$ ({\it Collision entropy}): 
When $q=2$, one obtains
\begin{equation}
    S_2^{\rm R\acute{e}nyi} = - \log \sum_i(p_i)^2
    \label{eq:Collision_entropy}
\end{equation}
the collision entropy, that can be interpreted as follows. Suppose that we perform two independent trials generated by the same probability distribution (denoted as $p_i$). The probability that the two trials give rise to the same specific event, say, an event $i$, is $(p_i)^2$. Thus $\sum_i(p_i)^2$ corresponds to the probability of observing the same event irrespective of the type of event. $S_2^{\rm R\acute{e}nyi}$ in Eq.~(\ref{eq:Collision_entropy}) is (the minus of) the logarithm of such probability. Alternatively, one can think about a joint system composed of two identical but independent {\it replicas} following the same probability distribution. $S_2^{\rm R\acute{e}nyi}$ corresponds to the logarithmic of the probability that the two replicas give rise to the same event. 
The above interpretations can be easily generalized to higher (integer) values of $q$: $S_q^{\rm R\acute{e}nyi}$ in Eq.~(\ref{eq:def_Renyi}) corresponds to the logarithm of the probability of observing the same event for all of $q$ replicas (which are identical yet independent). 
The notion of replicas arises therefore naturally in the Rényi entropy, and makes it easier to measure in simulations and experiments than the Shannon entropy. 
Indeed $\sum_i(p_i)^q$ in the Rényi entropy can be evaluated by making $q$ copies of the system, while the logarithmic form, $\sum_i p_i \log p_i$ in the Shannon entropy is not straightforward to measure.
This is one of the main reasons why it has been used in various contexts, in particular dynamical systems~\cite{grassberger1983estimation,halsey1986fractal} and quantum many-body systems~\cite{Islam2015,kaufman2016quantum}. In the next sections we discuss further this connection with replicas (see also Ref.~\cite{kurchan2010order} for a detailed discussion).
%\MO{I agree.}
%This connection will be discussed further in the next sections.
%\NJ{can mention here that this is exactly how Rényi 2 was measured in a quantum experiment \cite{Islam2015}}

%We note that $\sum_i(p_i)^q$ in Eq.~(\ref{eq:def_Renyi}) with $q \geq 2$ is relatively easier to measure than $\sum_i p_i \log p_i$ in Eq.~(\ref{eq:def_Shannon}) in simulations as well as real experiments, which is one of the main reasons why the Rényi entropy has been used in various physics domains, such as dynamical systems~\cite{grassberger1983estimation} and quantum many-body systems~\cite{Islam2015}.

\vspace{0.5cm}
\noindent
iv) $q \to \infty$ ({\it Min-entropy}): 

When $q$ is very large, as one can expect from Fig.~\ref{fig:p_q}, the event with the highest probability dominates the summation, $\sum_i (p_i)^q$, in Eq.~(\ref{eq:def_Renyi}), i.e., $\sum_i (p_i)^q \simeq (\max_i \{ p_i \})^q$.
Thus, one gets
\begin{equation}
    S_{\infty}^{\rm R\acute{e}nyi} = - \log \max_i \{ p_i \} = \min_i \{ - \log p_i \} .
    \label{eq:Min_entropy}
\end{equation}
This is called the min-entropy in a sense that only the event with the minimum of $I_i = - \log p_i$ (or the highest probabiliy) is taken into account.

While $S_0^{\rm R\acute{e}nyi}$ in Eq.~(\ref{eq:Max_entropy}) weights all the events with $p_i>0$ on an equal footing in the summation $\sum_i (p_i)^q$, $S_{\infty}^{\rm R\acute{e}nyi}$ in Eq.~(\ref{eq:Min_entropy}) instead weights only the event with the largest probability. The Shannon entropy $S^{\rm Shannon}$ in Eq.~(\ref{eq:def_Shannon}) weights the events in an average manner, located in the middle of the two extreme limits, $q\to 0$ and $q \to \infty$.
As one can expect from Fig.~\ref{fig:p_q}, in general, a larger value of $q$ in $S_q^{\rm R\acute{e}nyi}$ tends to discriminate or highlight events with larger probability, while a smaller value of $q$ tends to take into account events with finite probabilities on rather equal manner. Thus varying $q$ from the Shannon entropy limit ($q \to 1$) corresponds to biasing (or unbiasing) the original probability distribution.
This view is naturally related to the large deviation theory~\cite{touchette2009large,jack2020ergodicity}, as we will discuss below.

To see the $q$-dependence of $S_q^{\rm R\acute{e}nyi}$ more systematically,
we compute its derivative:
\begin{equation}
    \frac{\partial S_q^{\rm R\acute{e}nyi}}{\partial q} = - \frac{\sum_i \tilde p_i^q \log (\tilde p_i^q/p_i)}{(1-q)^2} = - \frac{D_{\rm KL}(\tilde {\bf p}^q || {\bf p})}{(1-q)^2} \leq 0 ,
\end{equation}
where $\tilde {\bf p}^q=(\tilde p_1^q, \tilde p_2^q, \cdots, \tilde p_\Omega^q)$ is a normalized probability distribution whose component is given by $\tilde p_i^q = (p_i)^q/\sum_j (p_j)^q$ (and hence $\sum_i \tilde p_i^q=1$), and $D_{\rm KL}$ is the (standard) Kullback-Leibler divergence~\cite{cover1999elements} which is non-negative.
Therefore, $S_q^{\rm R\acute{e}nyi}$ is a non-increasing function of $q$, satisfying $S_0^{\rm R\acute{e}nyi} \geq S_q^{\rm R\acute{e}nyi} \geq S_{\infty}^{\rm R\acute{e}nyi}$, and $S_q^{\rm R\acute{e}nyi}$ is bounded by $S_{\infty}^{\rm R\acute{e}nyi}$ from below.
Moreover, when $q>1$ one can also derive an upper bound of $S_q^{\rm R\acute{e}nyi}$ by $S_{\infty}^{\rm R\acute{e}nyi}$.
Since $\sum_i(p_i)^q \geq \max_i \{ (p_i)^q \} = (\max_i \{ p_i \})^q$, and using Eqs.~(\ref{eq:def_Renyi}) and (\ref{eq:Min_entropy}), we get $(q-1) S_q^{\rm R\acute{e}nyi} \leq q S_{\infty}^{\rm R\acute{e}nyi}$. Thus, when $q>1$,
\begin{equation}
     S_{\infty}^{\rm R\acute{e}nyi} \leq S_q^{\rm R\acute{e}nyi} \leq \frac{q}{q-1} S_{\infty}^{\rm R\acute{e}nyi} .
     \label{eq:inequality_Renyi}
\end{equation}
Several other inequalities are summarized in Ref.~\cite{zyczkowski2003Renyi}.

%The Rényi entropy with various values of $q$ has been widely used in physics, such as an estimate of entanglement entropy in quantum many-body systems (e.g., $q=2$)~\cite{kaufman2016quantum}, non-equilibrium statistical mechanics ($q=1/2$ and $q \to \infty$)~\cite{dahlsten2011inadequacy,halpern2015introducing}, multi-fractal analysis~\cite{jizba2004world}, characterization of chaos ($q=2$)~\cite{grassberger1983estimation}, as well as estimation of the Shannon entropy from the Rényi entropy~\cite{zyczkowski2003Renyi}.\NJ{I think this is a bit redundant with was has been already said before, can we move it to the introduction ? Also, it could be worth giving more detail about the interpretation of $q=1/2$ if you mention it ?}\MO{Yes I agree. Could you edit the text accordingly? I do not have a specific interpretation for $q=1/2$. It was just to show many different $q$ values are used in physics.}
% One of the main goals of this paper is to provide concise interpretations of the Rényi entropy with index $q$ in physical systems.
%\NJ{i'm not sure Rényi 2 is generally denoted as "entanglement entropy" in the field; I think in that paper, purity (Rényi 2) is used as a proxy to quantify entanglement}\MO{"an estimate" is ok?}

\section{Rényi entropy for the Gibbs-Boltzmann distribution}

We consider the Rényi entropy for the Gibbs-Boltmann distribution at a temperature $T=\beta^{-1}$, given by
\begin{eqnarray}
p_i &=& \frac{e^{-\beta E_i}}{Z(T)}, \label{eq:Boltmann} \\
Z(T) &=& \sum_i e^{-\beta E_i}, \label{eq:Z} \\
-\beta F(T) &=& \log Z(T), \label{eq:F}
\end{eqnarray}
where $Z(T)$ and $F(T)$ are the partition function and free energy, respectively.
The index $i$ specifies a configuration, and $E_i$ is the corresponding energy. 
In this paper we focus on the standard Gibbs-Boltzmann distribution~\footnote{Note that maximization of various generalized entropies (including the Rényi entropy) has been discussed~\cite{tsallis1988possible,parvan2005extensive,bakiev2020certain,beck2009generalised}, which amounts to generalized statistical mechanics with generalized probability distribution.} in Eq.~(\ref{eq:Boltmann}) obtained by maximizing the Shannon entropy under the constraint on the average energy, $\langle \hat E \rangle = \sum_i p_i E_i$~\cite{jaynes1957information},
where $\langle (\cdots) \rangle=\sum_i p_i (\cdots)$ is the expectation value given by the linear average (arithmetic mean) and $\hat X$ represents a stochastic variable of $X$ in statistical averages.
%We do not discuss along this line of thought. We discuss the Rényi entropy within the standard statistical mechanics based on the Gibbs-Boltzmann distribution.
%we do not discuss further this line of thought, but focus on standard statistical %mechanics based on the Gibbs-Boltzmann distribution.

\vspace{0.5cm}
{\it Free energy:}
In the statistical mechanics setting above, with Eqs.~(\ref{eq:Boltmann}), (\ref{eq:Z}), and (\ref{eq:F}), the Rényi entropy $S_q^{\rm R\acute{e}nyi}$ in Eq.~(\ref{eq:def_Renyi}) becomes~\cite{baez2022Renyi,fuentes2022Renyi}
\begin{eqnarray}
S_q^{\rm R\acute{e}nyi}(T) &=& \frac{1}{1-q} \log \frac{Z(T/q)}{\left(Z(T)\right)^q} 
\label{eq:representation0}    
\\
&=& - \frac{F(T/q)-F(T)}{T/q-T} .
\label{eq:representation1}    
\end{eqnarray}
This equation tells us that $S_q^{\rm R\acute{e}nyi}(T)$ has a physical meaning of free energy difference between a state at $T$ and a state at $T/q$ (which is a lower temperature when $q>1$). 
Thus, $S_q^{\rm R\acute{e}nyi}(T)$ can be obtained by measuring the equilibrium free energy at $T$ and $T/q$.
Alternatively, measuring $S_q^{\rm R\acute{e}nyi}(T)$ at a given temperature $T$ allows us to access the free energy at lower temperatures~\cite{sakaguchi1989Renyi}.
When $q \to 1$, one can easily check that the Rényi entropy corresponds to the Shannon entropy (or the thermodynamic entropy): $S_1^{\rm R\acute{e}nyi}(T) = - \partial F(T)/\partial T = S^{\rm Shannon}(T)$.
Equation~(\ref{eq:representation1}) is rewritten as the integral of the Shannon entropy from $T$ to $T/q$~\cite{johnson2019physical}:
\begin{equation}
    S_q^{\rm R\acute{e}nyi}(T) = \frac{1}{T/q-T} \int_T^{T/q} d T' S^{\rm Shannon}(T') .
\end{equation}

\begin{figure}
\begin{center}
\includegraphics[width=0.8\linewidth]{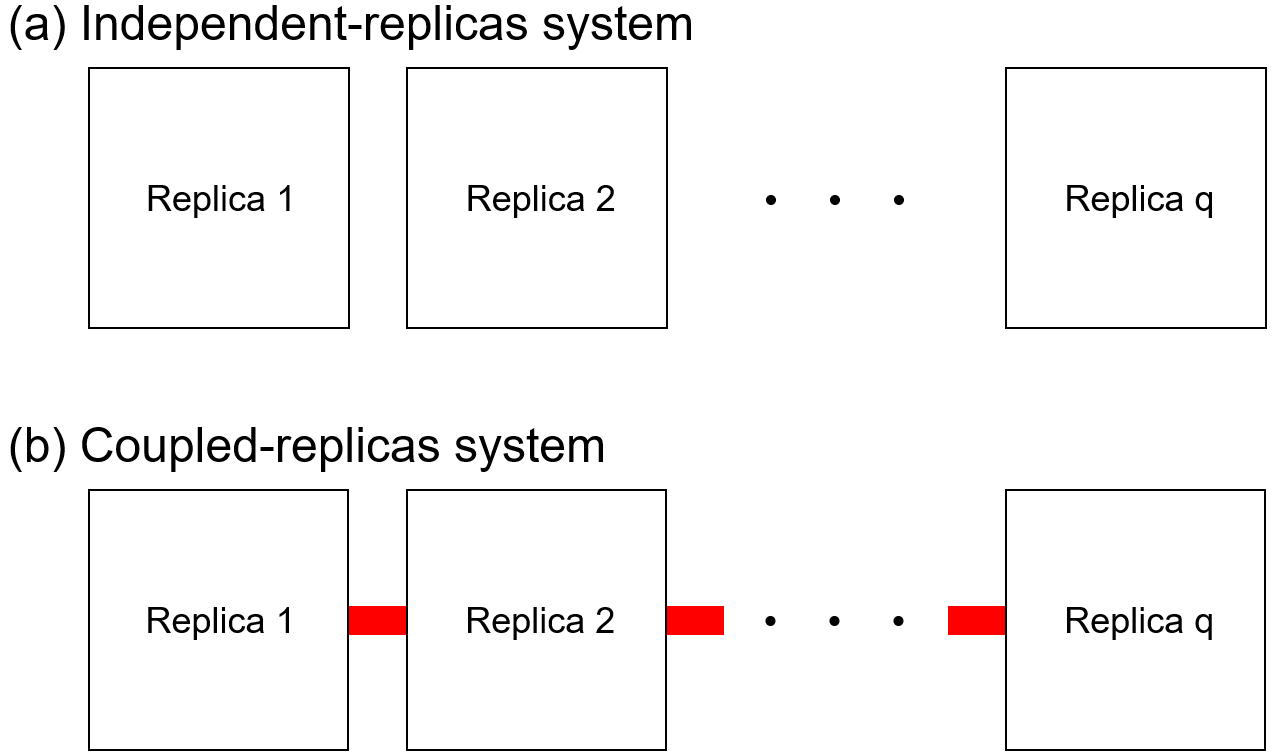}
\end{center}
\caption{Schematic plot for a system composed of $q$ replicas. (a): Independent-replicas system. (b): Coupled-replicas system. Interaction potentials in Eq.~(\ref{eq:external_potential}) are schematically represented as the horizontal (red) lines.}
\label{fig:schematic_replicas}
\end{figure}

\vspace{0.5cm}
{\it Replicas:}
We consider a different representation using the notion of $q$ replicas, which provides another physical interpretation of $S_q^{\rm R\acute{e}nyi}(T)$. When $q$ is integer with $q \geq 2$, Equation~(\ref{eq:representation0}) can be rewritten as~\cite{calabrese2004entanglement} 
\begin{equation}
    S_q^{\rm R\acute{e}nyi}(T) = \frac{1}{1-q} \log \frac{Z_{\rm couple}(T,q)}{Z_{\rm indep}(T,q)} ,
    \label{eq:S_q_indep_couple}
\end{equation}
where $Z_{\rm indep}(T,q)$ and $Z_{\rm couple}(T,q)$ are given by
\begin{eqnarray}
    Z_{\rm indep}(T,q) &=& \sum_{i_1, i_2, \cdots, i_q} e^{-\beta(E_{i_1} + E_{i_2} + \cdots + E_{i_q})} , \\
    Z_{\rm couple}(T,q) &=& \sum_i e^{-q \beta E_i} .
    \label{eq:Z_couple_1}
\end{eqnarray}
Thus $Z_{\rm indep}(T,q)$ is the partition function of the system composed of $q$ independent replicas that are not interacting with each other (independent-replicas system) as schematically shown in Fig.~\ref{fig:schematic_replicas} (a).
Instead, an interpretation of $Z_{\rm couple}(T,q)$ is as follows.
We consider an external potential $V^{\rm ext}_{i_1 i_2 \cdots i_q}$ associated with interaction between replicas that condense all $q$ replicas into the same configuration (see Fig.~\ref{fig:schematic_replicas} (b)).
$V^{\rm ext}_{i_1 i_2 \cdots i_q}$ can be written formally~\footnote{To formulate it more properly, we could introduce a regularization, e.g., replacing $\log \delta_{i_1 i_2}$ with $\log (\delta_{i_1 i_2} + \epsilon) -  \log (1 + \epsilon)$ with a small $\epsilon>0$. One could also design this potential in system-specific ways, such as overlap functions in disordered systems~\cite{mezard1987spin}.} as
\begin{equation}
    \beta V^{\rm ext}_{i_1 i_2 \cdots i_q} = - \log \delta_{i_1 i_2} - \log \delta_{i_2 i_3} - \cdots - \log \delta_{i_{q-1} i_q} ,    
    \label{eq:external_potential}
\end{equation}
where $\delta_{ij}$ is the Kronecker delta.
$Z_{\rm couple}(T,q)$ in Eq.~(\ref{eq:Z_couple_1}) can now be rewritten as the partition function of the system composed of $q$ replicas interacting with each other (coupled-replicas system) by $V^{\rm ext}_{i_1 i_2 \cdots i_q}$:
\begin{equation}
    Z_{\rm couple}(T,q) = \sum_{i_1, i_2, \cdots, i_q} e^{-\beta \left( E_{i_1} + E_{i_2} + \cdots + E_{i_q} + V^{\rm ext}_{i_1 i_2 \cdots i_q} \right)} .
\end{equation}
We note that the coupling between replicas is very strong, such that all different replicas condense in the same configuration~
\footnote{As an alternative to Eq.~(\ref{eq:external_potential}), one can also consider an external potential with a master-slave architecture, e.g., $i_1$ is the master that interacts with all slaves, $i_2, i_3, \cdots, i_q$, and there is no interaction between slaves. One can also consider fully connected interactions, i.e., all $q$ replicas interact with each other, akin to the replica method in disordered systems~\cite{charbonneau2023spin}.}.

We now define the corresponding free energies for the independent- and coupled-replicas systems:
\begin{eqnarray}
 -\beta F_{\rm indep}(T, q) &=& \log Z_{\rm indep}(T, q), \\
-\beta F_{\rm couple}(T, q) &=& \log Z_{\rm couple}(T, q) .
\end{eqnarray}
With these expressions, one can rewrite Eq.~(\ref{eq:S_q_indep_couple}) as
\begin{equation}
    S_q^{\rm R\acute{e}nyi}(T) = \frac{ \beta \left( F_{\rm couple}(T, q)-F_{\rm indep}(T, q) \right) }{q-1} .
    \label{eq:representation2}
\end{equation}
Thus, $S_q^{\rm R\acute{e}nyi}(T)$ is interpreted as a free energy difference between the coupled- and independent-replicas systems.

We then consider a work $W^{\rm ext}$ done on the system at a temperature $T$ to condense all $q$ replicas in a same configuration, realized by the external potential $V_{i_1 i_2\cdots i_q}^{\rm ext}$. In this setting, we obtain an inequality associated with the second law of thermodynamics, i.e., 
\begin{equation}
 W^{\rm ext} \geq \Delta F= F_{\rm couple}(T, q)-F_{\rm indep}(T, q) .   
 \label{eq:work_inequality}
\end{equation} 
The equality holds if the work is performed quasi-statically, while $W^{\rm ext}$ can be larger than the free energy difference $\Delta F$ in non-equilibrium processes.
Therefore, with Eqs.~(\ref{eq:work_inequality}) and (\ref{eq:representation2}), the Rényi entropy provides a lower bound on the external work done on the system to condense $q$-independent replicas into a same configuration:
\begin{equation}
    W^{\rm ext} \geq T(q-1) S_q^{\rm R\acute{e}nyi}(T) .
    \label{eq:work}
\end{equation}
This is another physical interpretation of the Rényi entropy, using the notion of replicas and work.

Since $S_q^{\rm R\acute{e}nyi}(T)$ in Eq.~(\ref{eq:representation2}) is represented as a free energy difference, it can be evaluated by Monte-Carlo simulaitons with various techniques~\cite{iaconis2013detecting,alba2016classical}, such as thermodynamic integration~\cite{frenkel2023understanding} and free energy perturbation~\cite{zwanzig1954high}. 
For example, the free energy difference in Eq.~(\ref{eq:representation2}) can be written as a statistical average of the external potential $V^{\rm ext}_{i_1 i_2 \cdots i_q}$: 
\begin{equation}
    \beta \left( F_{\rm couple}(T, q)-F_{\rm indep}(T, q) \right) = - \log\\ \left \langle e^{-\beta \hat V^{\rm ext}} \right\rangle_{\rm indep} ,
\end{equation}
where
\begin{equation}
    \left\langle e^{-\beta \hat V^{\rm ext}} \right\rangle_{\rm indep} = \sum_{i_1, i_2, \cdots, i_q} \frac{e^{-\beta (E_{i_1}+E_{i_2} + \cdots + E_{i_q})}}{Z_{\rm indep}(T, q)}e^{-\beta V^{\rm ext}_{i_1 i_2 \cdots i_q}} 
\end{equation}
is a statistical average performed over the independent-replicas system.
%\MO{In this paper $\string^$ is used to represent stochastic variables in statistical averages.}

Since the external potential $V^{\rm ext}_{i_1 i_2 \cdots i_q}$ is quite strong, standard equilibrium sampling methods might not be suitable. In such a case, the Jarzynski method~\cite{jarzynski2004nonequilibrium}, which monitors the work along a non-equilibrium trajectory and average it over many paths, would be effective, as demonstrated in Refs.~\cite{alba2017out,d2020entanglement,zhao2022measuring}.

The representation in Eq.~(\ref{eq:representation1}) allows us to access the free energy $F(T/q)$ at a lower temperature $T/q$ (when $q>1$) using $S_q^{\rm R\acute{e}nyi}(T)$ at a temperature $T$. 
In another representation in Eq.~(\ref{eq:representation2}), $S_q^{\rm R\acute{e}nyi}(T)$ is connected to the free energy of the coupled-replicas system with $q$ replicas, $F_{\rm couple}(T,q)$. By combining these two representations, one can translate $F(T/q)$ at $T/q$ into $F_{\rm couple}(T,q)$ at $T$. It would be interesting to exploit this relationship in order to develop efficient free energy computation methods at lower temperatures.

%\NJ{\textit{maybe add that free energy computation at low temperatures is a challenge in the glass community ? This might not be obvious to non-glass people}}
%\MO{Free energy computations at lower temperatures are challenging in many fields, soft matter, biophysics, biochemistry, etc. So as it is fine.}

\vspace{0.3cm}
{\it Rényi energy:}
The arguments under Eq.~(\ref{eq:representation1}) show that the Rényi entropy with $q>1$ at a temperature $T$ contains information about lower temperature $T/q$, suggesting that the Rényi entropy is related to lower energy states. We now discuss this aspect in detail.
First, $S_q^{\rm R\acute{e}nyi}(T)$ can be rewritten as
\begin{eqnarray}
    S_q^{\rm R\acute{e}nyi}(T) &=& \frac{1}{1-q} \log \sum_i p_i (p_i)^{q-1} \nonumber \\
    &=& - \frac{1}{q-1} \log \left\langle e^{-(q-1)\beta \hat E} \right\rangle - \beta F(T) .
    \label{eq:Renyi_calculation}
\end{eqnarray}
We define a generalized mean of the energy, the Rényi energy $\beta E_q^{\rm R\acute{e}nyi}(T)$, defined analogously to the entropy in Eq.~(\ref{eq:mean_general}),
\begin{eqnarray}
    \beta E_q^{\rm R\acute{e}nyi}(T) &=& f^{-1}\left( \sum_i p_i f(\beta E_i) \right) \nonumber \\
    &=& - \frac{1}{q-1} \log \left\langle e^{-(q-1)\beta \hat E} \right\rangle ,\label{eq:Def_generalized_energy}
\end{eqnarray}
where $f(x)=e^{-(q-1)x}$.
Thus, Eq.~(\ref{eq:Renyi_calculation}) becomes
\begin{equation}
    S_q^{\rm R\acute{e}nyi}(T) = \beta E_q^{\rm R\acute{e}nyi}(T) - \beta F(T).
    \label{eq:Renyi_thermodynamics}
\end{equation}
This is a generalization of the following usual expression for the thermodynamic (Shannon) entropy,
\begin{equation}
    S^{\rm Shannon}(T) = \beta \langle \hat E \rangle - \beta F(T) .    \label{eq:shannon_thermodynamics}
\end{equation}

\begin{figure}
\begin{center}
\includegraphics[width=0.85\linewidth]{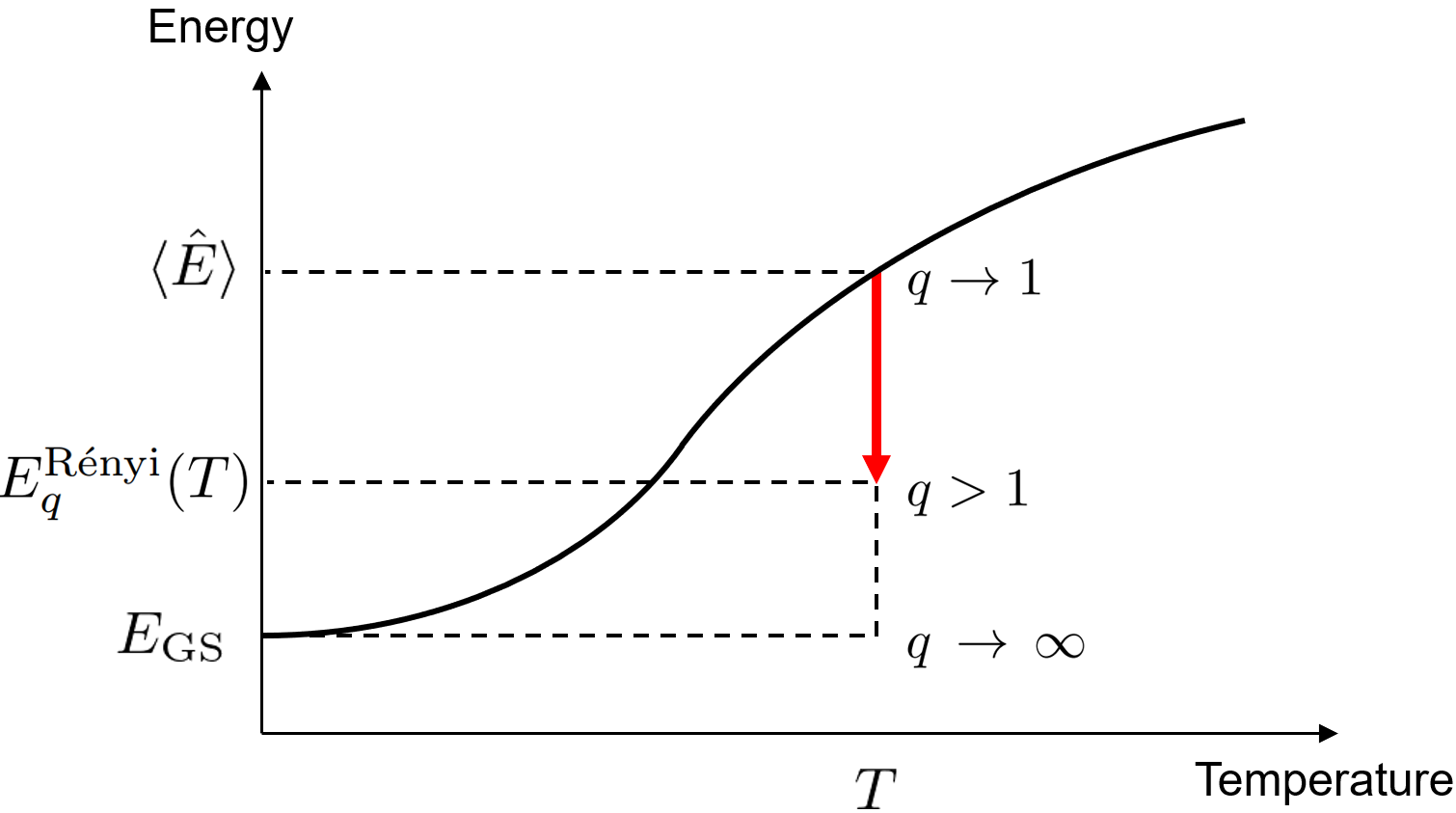}    
\end{center}
\caption{Schematic plot for the energy versus temperature. The solid curve is the mean equilibrium energy $\langle \hat E \rangle$. The Rényi energy $E_q^{\rm R\acute{e}nyi}(T)$ goes down along the vertical axis when $q$ is increased at a constant temperature $T$. The $q \to 1$ and $q \to \infty$ limits correspond to $\langle \hat E \rangle$ and $E_{\rm GS}$ (the ground state energy), respectively. 
}
\label{fig:schematic_enery_vs_temperature}
\end{figure}

$E_q^{\rm R\acute{e}nyi}(T)$ has the following properties.
As expected, $\lim_{q \to 1} E_q^{\rm R\acute{e}nyi}(T)= \langle \hat E \rangle$.
When $q \to \infty$, since
$\lim_{q \to \infty} E_q^{\rm R\acute{e}nyi}(T) = T S_\infty^{\rm R\acute{e}nyi}(T) + F(T) = T \min_i \{ - \log p_i\} + F(T)$ and $- \log p_i = \beta E_i - \beta F(T)$, we obtain
\begin{equation}
    \lim_{q \to \infty} E_q^{\rm R\acute{e}nyi}(T) = \min_i \{ E_i \} = E_{\rm GS},
    \label{eq:ground_state}
\end{equation}
where $E_{\rm GS}$ is the ground state energy of the system.
The relationships between $E_q^{\rm R\acute{e}nyi}(T)$, $\langle \hat E \rangle$, and $E_{\rm GS}$ are schematically shown in Fig.~\ref{fig:schematic_enery_vs_temperature}.
In general, $E_q^{\rm R\acute{e}nyi}(T)$ goes down along the vertical axis toward $E_{\rm GS}$ (at a constant $T$) when $q$ is increased, which implies that increasing $q$ allows us to sample configurations associated with a lower energy while keeping a constant temperature $T$.

Using Eqs.~(\ref{eq:Renyi_thermodynamics}) and (\ref{eq:shannon_thermodynamics}), the energy difference between $\langle \hat E \rangle$ and $E_q^{\rm R\acute{e}nyi}(T)$ (the vertical arrow in Fig.~\ref{fig:schematic_enery_vs_temperature}) is given by the difference between $S^{\rm Shannon}(T)$ and $S_q^{\rm R\acute{e}nyi}(T)$ as follows:
\begin{equation}
    \langle \hat E \rangle - E_q^{\rm R\acute{e}nyi}(T) = T \left( S^{\rm Shannon}(T)-S_q^{\rm R\acute{e}nyi}(T) \right) .
    \label{eq:mean_energy_minus_ground_state}
\end{equation}

Equations~(\ref{eq:Renyi_thermodynamics}) and (\ref{eq:ground_state}) also allow us to estimate the ground state energy $E_{\rm GS}$ using equilibrium values of $S_q^{\rm R\acute{e}nyi}(T)$ (with a large $q$) and $F(T)$ at a finite temperature $T$ as
\begin{equation}
    E_{\rm GS}=TS_\infty^{\rm R\acute{e}nyi}(T) + F(T) .
\end{equation}
The estimation of $E_{\rm GS}$ is a non-trivial task in some cases, such as disordered systems. Thus, the Rényi entropy might shed light on this problem from a different angle. We note however that the evaluation of the Rényi entropy for very large $q$ would still be hard. 
In such a case, we can obtain a lower bound of $E_{\rm GS}$ by $S_q^{\rm R\acute{e}nyi}(T)$ at a finite $q$ and using the inequality in Eq.~(\ref{eq:inequality_Renyi}), $(1-q^{-1}) S_q^{\rm R\acute{e}nyi} \leq S_{\infty}^{\rm R\acute{e}nyi}$, 
\begin{equation}
   E_{\rm GS} \geq T(1-q^{-1}) S_q^{\rm R\acute{e}nyi}(T)+ F(T)  .
\end{equation}
One can see that the bound will be improved with increasing $q$.

\vspace{0.3cm}
{\it Large deviation:}
So far we have discussed that increasing $q>1$ can allow us to access lower energy configurations that can be considered rare or atypical at a fixed temperature $T=1/\beta$. We now argue this aspect in detail based on the large deviation theory~\cite{touchette2009large,jack2020ergodicity,mora2016Renyi}.
Suppose that the probability distribution of the intensive energy variable, $\hat e=\hat E/N$, follows the large deviation principle (as expected in many physical systems), 
\begin{equation}
    P_N(e) \approx e^{-N I(e)} ,
    \label{eq:large_deviation}
\end{equation}
where $I(e)$ is the rate function, and $N$ is the number of constituent elements (e.g., particles, spins).
When $N \to \infty$, the mean value of the energy $\hat e$ converges to $\langle \hat E \rangle/N$, with probability one (law of large numbers), which corresponds to the minimization of $I(e)$ at $I(e)=0$. The variance of $\hat e$ would converge to zero with $N^{-1}$ (central limit theorem).
The rate function $I(e)$ contains further information about large deviations~\cite{touchette2009large}. Following Refs.~\cite{kurchan2010order,mora2016Renyi}, we will show that the Rényi entropy contains the same information as $I(e)$.

Most importantly, as expected from the functional form in Eq.~(\ref{eq:Def_generalized_energy}), the Rényi energy $\beta E_q^{\rm R\acute{e}nyi}(T)$ and hence $S_q^{\rm R\acute{e}nyi}(T)$ (through Eq.~(\ref{eq:Renyi_thermodynamics})) are directly related to a cumulant generating function $\lambda_q(T)$ which is defined as 
\begin{eqnarray}
    \lambda_q(T) &=& 
    -\frac{1}{N} \log \left\langle e^{-(q-1)\beta \hat E} \right\rangle \nonumber \\
    &=&
    \frac{(q-1)}{N}\beta E_q^{\rm R\acute{e}nyi}(T) \nonumber \\
    &=& \frac{(q-1)}{N} \left( S_q^{\rm R\acute{e}nyi}(T) + \beta F(T) \right) .
    \label{eq:def_generating_function}    
\end{eqnarray}
The large deviation theory~\cite{touchette2009large} shows that $\lambda_q(T)$ provides us with $I(e)$ by the Legendre transform, 
\begin{equation}
    I(e) = \min_q \left\{ \lambda_q(T) - (q-1) \beta e \right\} .
\end{equation}
Thus, $\beta E_q^{\rm R\acute{e}nyi}(T)$ and $S_q^{\rm R\acute{e}nyi}(T)$ in Eq.~(\ref{eq:def_generating_function}) contain the whole information about the large deviation in Eq.~(\ref{eq:large_deviation}), giving another interpretation of the Rényi entropy.

With Eqs.~(\ref{eq:representation2}), (\ref{eq:def_generating_function}), and $F_{\rm indep}(T,q)=qF(T)$, one can also show that 
\begin{equation}
    \lambda_q(T) = \frac{\beta \left( F_{\rm couple}(T,q)-F(T) \right)}{N} ,
\end{equation}
which translates the cumulant generating function $\lambda_q(T)$ into the replica terminology.

\section{Discussion and Conclusions}
The Rényi entropy is a one-parameter generalization of the Shannon entropy obtained by considering a generalized mean of the information content that preserves additivity (extensivity) for independent events, a fundamental aspect in information theory and a natural characteristic feature in many physical systems. Besides defining the Rényi entropy, statistical mechanics using this generalized expectation value allow to investigate large deviation, and possibly replica theory in a formally elegant manner~\cite{morales2023thermodynamics}. 

The goal of this paper is to give the readers an overview of the physical interpretations of the Rényi entropy. Thus, we did not discuss carefully the non-analyticity of $S_q^{\rm R\acute{e}nyi}$ in $q$, which might be an issue in some cases, e.g., phase transitions. This would be related to the cumulant generating function $\lambda_q$ in the large deviation theory being not differentiable~\cite{touchette2009large,jack2020ergodicity}. 
%Besides, we did not discuss the analytic continuation of the Rényi entropy in the replica parameter $q$ from integer
%to non-integer values. 
%\NJ{\textit{I would not phrase it like this because it makes the reader think we implicitly considered integer $q$, whereas we do so only when we give physical interpretation using replicas. As Eric noted, eq (15) makes sense for any $q$.}}
%\MO{The Rényi entropy can be defined in non-negative real numbers in general.   
%Yet, if one stands on the replica interpretation or measurements using replicas (with integer numbers), one may need to care about analytical continuation, which would be another delicate problem.} 
Similar issues were studied in the context of mean-field spin glasses ~\cite{kondor1983parisi,ogure2004exact,parisi2008large,nakajima2008large,pastore2019large}. 

We also restricted to systems with discrete events, such as spins. Extending to continuous systems, such as classical particle systems, is seemingly straightforward. Yet it involves additional diverging terms (e.g., the logarithm of the bin size for discretization) in the same way as the Shannon entropy~\cite{tabass2016renyi}, which requires careful analysis for the system under investigation.

%\NJ{\textit{maybe we could simply mention the analytic continuation in the intro, as an additional motivation for considering Rényi to get Shannon, and stress under (4) that non-integer $q$ makes perfect sense.}}

\acknowledgments

We thank Eric Bertin, David Horvath, and Jorge Kurchan for insightful discussions.
This work has been supported by MIAI@Grenoble Alpes, (ANR-19-P3IA-0003).

\bibliographystyle{eplbib} 

\bibliography{Renyi.bib}

\end{document}